\documentclass[11pt]{article}
\usepackage{moriond,epsfig}

\def\Journal#1#2#3#4{{#1} {\bf #2}, #3 (#4)}

\def\PLB{{\em Phys. Lett.}  B}

\def\EPJC{{\em Eur. Phys. J.} C}

\def\be{\begin{equation}}
\def\ee{\end{equation}}
\def\bea{\begin{eqnarray}}
\def\eea{\end{eqnarray}}

\begin{document}
\vspace*{4cm}
\title{Higgs Searches at LEP}

\author{ Arnulf Quadt }

\address{Physikalisches Institut, Universit\"at Bonn,\\ Nu{\ss}allee 12,\\ 53115 Bonn, Germany}

\maketitle\abstracts{
The four LEP experiments ALEPH, DELPHI, L3 and OPAL have performed
decay-mode independent searches for scalar bosons, flavour-independent
Higgs boson searches, searches for Higgs bosons in the Standard Model
and several Two-Higgs-Doublet models and searches for doubly charged
Higgs bosons at centre-of-mass energies up to 209 GeV. The results
obtained and mostly combined by the LEP Higgs working group are
discussed.}

\section{Introduction}
During the last years of running of the $e^+ e^-$ collider LEP,
searches for Higgs bosons and hence for the origin of electroweak
symmetry breaking have been of highest interest. The search for the
MSSM and in particular for the Standard Model (SM) Higgs bosons have
been pursued by the four LEP experiments and the results have been
combined by the LEP Higgs Working Group for optimal search
sensitivity. This article describes the present status of the search
for Higgs bosons at LEP, mostly results as combined by the LEP Higgs
working group, or analyses prepared for future combination by single
experiments.

Most Higgs searches at LEP were limited in their sensitivity by the
provided centre-of-mass energy. Hence they profited significantly from
the outstanding performance of the LEP machine experts, who managed to
push the centre-of-mass energies to extreme values, allowing the
experiments each to record about $700 \;\rm pb^{-1}$ of data above
$\sqrt{s} \approx 160\; \rm GeV$, $450 \;\rm pb^{-1}$ of which above
192 GeV and up to 209 GeV.

The data included in the analyses presented here have been subject to
at least one additional turn of calibration after data taking. In
particular the jet energy scales and the precision tracking and
b-tagging are highly sensitive to this and therefore reached in most
cases almost final precision.

The different Higgs boson searches in this presentation are ordered in
terms of increasing model dependence, starting from a fully decay-mode
independent search for scalar bosons, the search for purely
hadronically decaying Higgs bosons (`flavour-independent search') and
the Standard Model Higgs boson followed by various searches in
different types of 2-Higgs Doublet Models and their interpretations,
in particular in the MSSM model, and ending with the search for doubly
charged Higgs bosons.

\section{Decay-Mode Independent Higgs Search}
These analyses \cite{opal_pn495} represent topological searches for
new neutral scalar bosons $S^0$ with a minimum number of model
assumptions. The new bosons are only assumed to be produced in
association with a $Z^0$ boson via the Bjorken process $e^+ e^-
\rightarrow S^0 Z^0$, where $S^0$ denotes, depending on the context,
any new scalar neutral boson.
The analyses are based on studies of the recoil mass spectrum of $Z^0
\rightarrow e^+e^-$ and $\mu^+ \mu^-$ events and on a search for
$S^0Z^0$ events with $S^0 \rightarrow e^+e^-$ or photons and $Z^0
\rightarrow \nu\bar{\nu}$. They are sensitive to all decays of the
$S^0$ into an arbitrary combination of hadrons, leptons, photons and
invisible particles, and to the case of a long-lived $S^0$ leaving the
detector without interacting. The analyses are applied to the full LEP
1 $Z^0$ resonance data ($115.4\;\rm pb^{-1}$ at $\sqrt{s} = 91.2\;\rm
GeV$) and the $662.4\;\rm pb^{-1}$ of LEP 2 data collected at
centre-of-mass energies in the range of 183 to \hbox{209 GeV}.

\begin{figure}[ht]
\psfig{figure=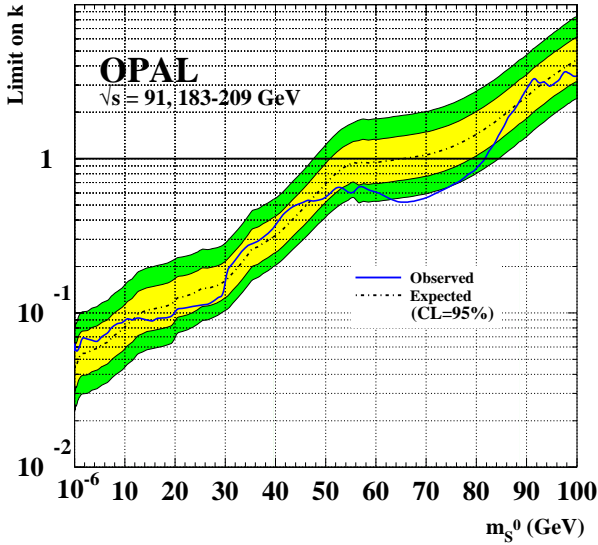,width=0.49\textwidth}
\hspace*{0.02\textwidth}
\parbox[]{0.47\textwidth}{
\vspace*{-3.50cm}
\caption[t]{
         \label{fig:modindep1} Exclusion limit for the scale factor
         $k$ on the cross-section for the production of a new scalar
         boson in the Higgs-Strahlung process (solid line), together
         with the expected median for background-only experiments
         (dot-dashed) and the 68\% and 95\% probability intervals
         shows as light and dark shaded bands.}}
\end{figure}

The results (figure~\ref{fig:modindep1}) are presented in terms of
limits on the scaling factor $k$, which relates the $S^0Z^0$
production cross-section to the SM cross-section for the
Higgs-Strahlung process via:
\begin{eqnarray}
\sigma_{S^0Z^0} = k \cdot \sigma_{H^0_{SM}}(m_{H^0_{SM}} = m_{S^0})
\end{eqnarray}
where it is assumed that $k$ does not depend on the centre-of-mass
energy for any given $m_{S^0}$. 

Values for $k \ge 0.1$ are excluded for $m_{S^0}$ below 19 GeV,
whereas $k = 1$, i.e. the SM rate for Higgs boson production, is
excluded for $m_{S^0}$ up to 81 GeV, independent of the decay modes of
the $S^0$ boson. For masses well below the width of the $Z^0$,
i.e. $m_{S^0} \le 1\;\rm GeV$, the obtained limits remain constant at
the level of $k^{95}_{obs.} = 0.067$, and $k^{95}_{exp.} = 0.051$.

These results are also used to investigate and interprete two
scenarios with continuous mass distribution, due to one single broad
state (`stealthy Higgs scenario') or several states close in mass
(`uniform Higgs scenario') \cite{opal_pn495}. The results will also be
included in future model interpretations and combinations by the LEP
Higgs working group.

\section{Flavour-Independent Higgs Search \label{sec:flavindep}}
There are extensions of the Standard Model in which Higgs bosons have
suppressed couplings to b-quarks. This can occur for specific
parameters of the Two Higgs Doublet Model, or of the Minimal
SuperSymmetric Model, as well as for some composite models. Standard
Model searches would have a reduced sensitivity in such cases, because
of their strong reliance on the identification of the b-quark from
the Higgs boson decay to maximize the separation power. It is
important to cover such scenarios experimentally with dedicated
searches in which the information from the flavour of the quarks in
the Higgs boson decay is not exploited, so that the model dependence
of the final Higgs search result can be reduced.

All four LEP collaborations have pursued such {\it
flavour-independent} searches in the recent years, analysing the
four-jet ($q\bar{q}q\bar{q}$), missing energy ($q\bar{q}\nu\bar{\nu}$)
and leptonic ($q\bar{q}l^+l^-$) topologies. None has found evidence
for a signal. The LEP Higgs Working Group has combined the search
results for the $e^+e^- \rightarrow {Z^0}^* \rightarrow hZ^0$
production mechanism \cite{lhwg_flavour_indep}. They are presented in
terms of upper limits on the corresponding cross-section as a function
of the Higgs bosons mass, and of a lower limit on the mass, assuming a
production cross-sections equal to those in the Standard Model and the
Higgs boson decaying to 100\% into hadrons. Higgs mass assumptions
from 60 to 115 GeV were tested.  The analyses have been mostly
developed based on the corresponding Standard Model analyses, but
removing the $b$-tag and using test-mass dependent selections.

The combined observed and median expected limits are $\rm 112.9$ and
$\rm 113.0\;\rm GeV/c^2$, respectively. The confidence level for the
background-only hypothesis fluctuates around 0.5 as expected for the
background. The sensitivity for a 5-sigma discovery is reached for a
$107\;\rm GeV/c^2$ Higgs boson.

These results have been and will be used in further model
interpretations, in particular for models with reduced Higgs coupling
to b-quarks.

\section{Standard Model Higgs Search}
The Standard Model predicts one single neutral scalar Higgs boson. Its
mass is arbitrary, but confined between about 130 and 190 GeV from
self-consistency of the model. Indirect experimental constraints from
precision measurements of electroweak parameters yield a preferred
Higgs boson mass of $m_H = 88^{+53}_{-33}\;\rm GeV$, and the 95\%
confidence level upper bound on the mass is 196 GeV
\cite{smfit_lep01}.

At LEP energies, the SM Higgs boson is expected to be produced mainly
in association with a $Z$ boson through the Higgs-Strahlung process
$e^+ e^- \rightarrow HZ$. For masses in the vicinity of 115 GeV (the
kinematic limit for Higgs-Strahlung at $E_{CM} \approx 206\;\rm GeV$),
the SM Higgs boson is expected to decay mainly into $b\bar{b}$ quark
pairs (74\%) while decays to tau lepton pairs, $WW^*$, gluon pairs
($\approx 7\%$ each), and $c\bar{c}$ ($\approx 4\%$) are all less
important. The final-state topologies are determined by these decays
and by the decay of the associated $Z$ boson. The searches at LEP
encompass the four-jet final state $(H
\rightarrow b\bar{b})q\bar{q}$, the missing energy final state $(H
\rightarrow b\bar{b}) \nu \bar{\nu}$, the leptonic final state $(H
\rightarrow b\bar{b})l^+l^-$ where $l$ denotes an electron or muon,
and the tau lepton final states $(H \rightarrow b\bar{b})
\tau^+ \tau^-$ and $(H \rightarrow \tau^+ \tau^-)(Z \rightarrow
q\bar{q})$.

The four LEP experiments perform their analyses in these search
channels using combinations of preselections and likelihood or neural
network techniques. The statistical analysis and combination of the
data is done using the likelihood ratio $-2 \ln Q$ as test-statistic.
Details on the analyses and the statistical combination method can be
found in \cite{lhwg_sm} and references therein.

\begin{figure}[ht]
         \centerline{\psfig{figure=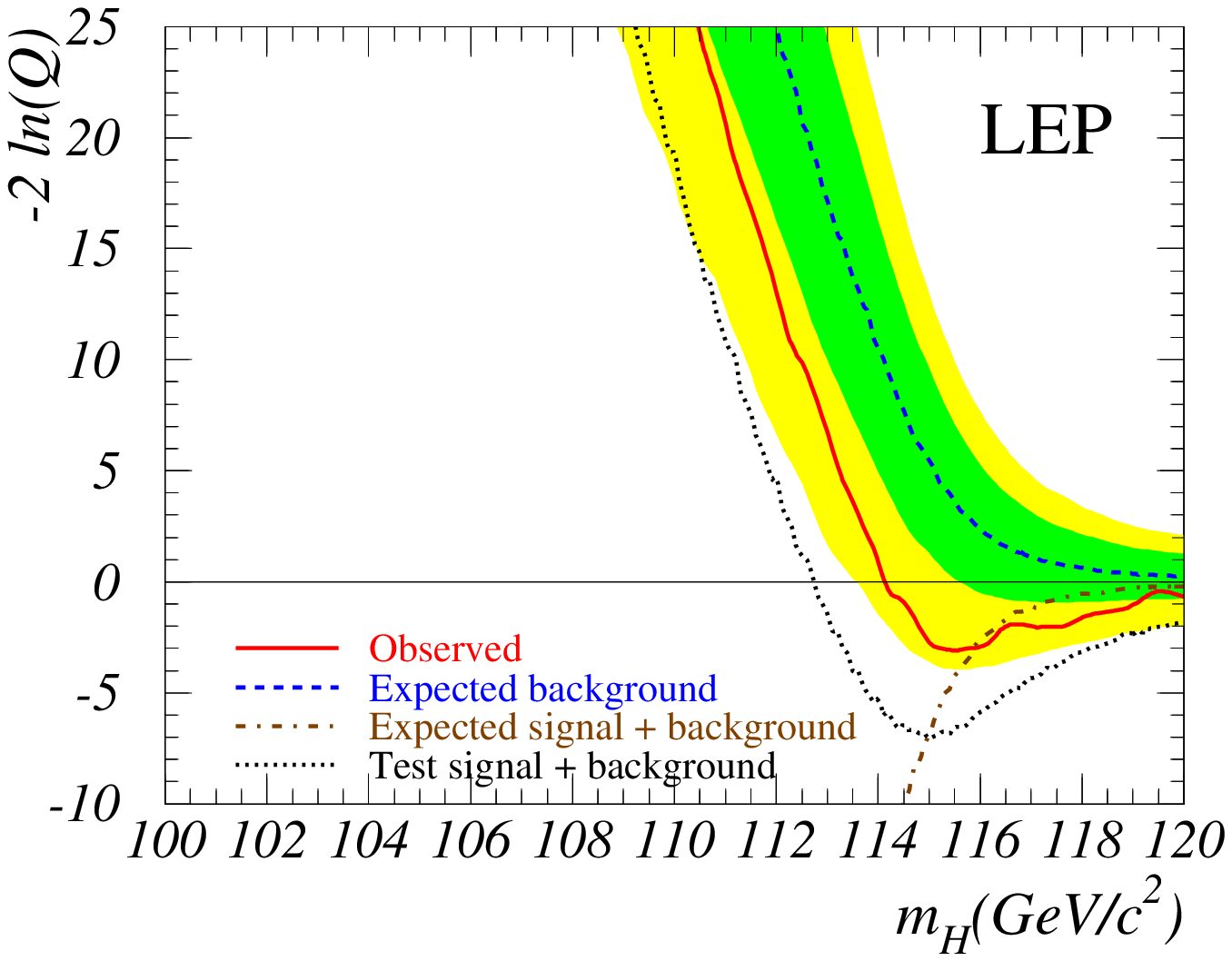,width=0.49\textwidth,clip=}
         \hfill
         \psfig{figure=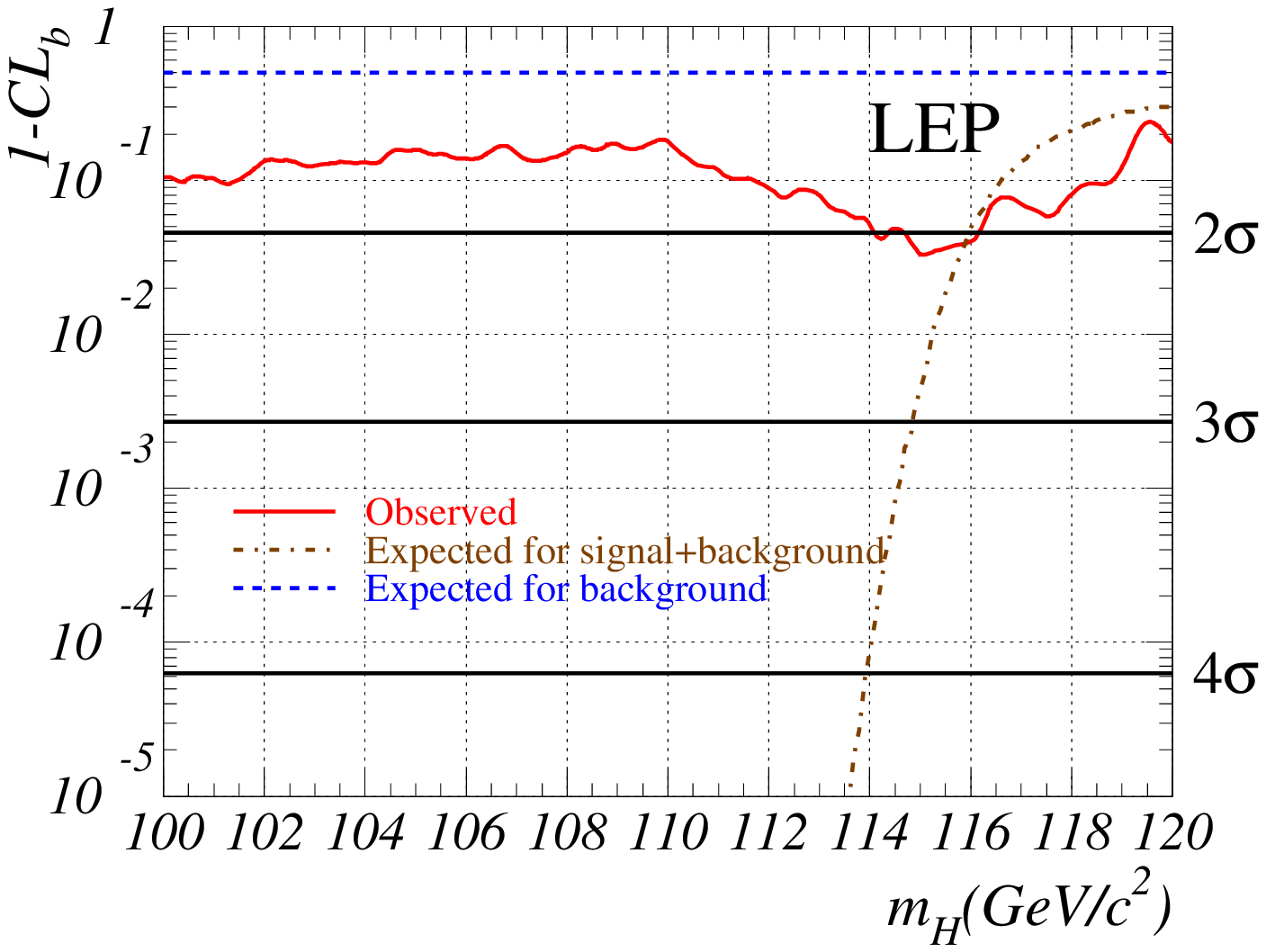,width=0.49\textwidth,clip=}}
         \caption{\label{fig:lhwg_sm_1} Left: Observed (solid line)
         and expected (dashed-dotted line) behaviour of the likelihood
         ratio $-2 \ln Q$ as a function of the test-mass
         $m_H$. Dark/light shaded bands around the median expected
         line correspond to the $\pm 1/ \pm 2$ standard deviation
         spreads; dotted line: result of a test where 115 GeV Higgs
         boson signal is added to background.  Right: The probability
         $1 - CL_b$ as a function of the test-mass $m_H$. Solid line:
         observation; dashed/dashed-dotted lines: expected probability
         for the background/signal+background hypotheses.}
\end{figure}
Figure~\ref{fig:lhwg_sm_1} shows the test-statistic $-2 \ln Q$ as a
function of the test-mass for the combination of LEP data. The
expected curves and their spreads are obtained by replacing the
observed data configuration by a large number of simulated event
configurations. The observed minimum in $-2 \ln Q$ at $m_H =
115.6\;\rm GeV$ indicates a deviation from the background
hypothesis. The persistent tail in the observation towards lower
test-masses where the observed curve stays below the prediction for
the background has been reproduced in a simulation where a 115 GeV
Higgs boson signal was injected in the background simulation and
propagated through the likelihood ratio calculation.  Studies on
contributions from individual experiments and final-state topologies
have shown that the signal-like behaviour in the vicinity of $m_H =
115\;\rm GeV$ mainly originates from the ALEPH data and is
concentrated in the four-jet channel. The DELPHI data and the combined
missing energy channel are rather background like.

Figure~\ref{fig:lhwg_sm_1} shows the confidence level $1 - CL_b$,
which is a measure of the compatibility of the observation with the
background hypothesis and calculated from $-2 \ln Q$, as a function of
the test-mass hypothesis. At $m_H = 115.6\;\rm GeV$, where the
likelihood ratio has its minimum, $1-CL_b = 0.034$, which corresponds
to about a two standard deviation excess in the LEP data.

The confidence level $CL_s = CL_{s+b}/CL_b$, which is a measure of the
compatibility with the signal+background hypothesis, is used to set
mass exclusion limits. The test-mass corresponding to $CL_s = 5\%$
defines the lower bound at the 95\% confidence level. The expected and
observed lower bounds obtained for the SM Higgs boson mass are 115.4
GeV and 114.1 GeV, respectively, at 95\% confidence level.

These results include the final L3 data \cite{l3_sm_final} and
preliminary data from ALEPH, DELPHI and OPAL. In the meantime ALEPH
has published their final analysis \cite{aleph_sm_final}. The final
publication by DELPHI and OPAL are imminent and will be followed by a
last and final combination of the LEP Higgs Working Group, expected
for the summer 2002.

\section{Two-Higgs Doublet Model Higgs Searches}
\subsection{Introduction}
Beyond the Standard Model, with only one physical neutral Higgs
scalar, it is important to study extended models containing more than
one physical Higgs boson. In particular, Two Higgs Doublet Models
(2HDM) are attractive extensions of the SM since they add new
phe\-no\-me\-na with the fewest new parameters: they satisfy the
constraints of $\rho
\approx 1$ and the absence of tree-level flavour changing neutral
currents, if the Higgs-fermion couplings are appropriately chosen. In
the context of 2HDMs the Higgs sector comprises five physics Higgs
bosons: two neutral CP-even scalars, $h^0$ and $H^0$ (with $m_h <
m_H$), one CP-odd scalar, $A^0$, and two charged scalars, $H^\pm$.  At
the centre-of-mass energies accessed by LEP, the $h^0$ and $A^0$
bosons are expected to be produced predominantly via two processes:
the Higgs-Strahlung process $e^+e^- \rightarrow h^0Z^0$ and the
associated-production $e^+e^- \rightarrow h^0A^0$. The cross-sections
for these two processes are related at tree level to the SM
cross-section by the following relations:
\begin{eqnarray}
e^+e^- \rightarrow h^0 Z^0 & : & \hspace*{3mm} \sigma_{hZ} = \sin^2(\beta - \alpha) \sigma_{HZ}^{SM} \label{eqn:2hdm_1}\\
e^+e^- \rightarrow h^0 A^0 & : & \hspace*{3mm} \sigma_{hA} = \cos^2(\beta - \alpha) \bar{\lambda} \sigma_{HZ}^{SM} \label{eqn:2hdm_2}
\end{eqnarray}
where $\tan \beta = v_2/v_1$ is the ratio of the the vacuum
expectation values, $\alpha$ is the mixing angle between $h$ and $H$
and $\bar{\lambda}$ is a kinematic term.

In the 2HDM (type II) the first Higgs doublet $(\Phi_1)$ couples only
to down-type fermions and the second Higgs doublet $(\Phi_2)$ couples
only to up-type fermions. In the type I model, quarks and leptons do
not couple to the first Higgs doublet, but couple to the second Higgs
doublet.

\subsection{2HDM Parameter Scan}
Using all available channels with and without b-tagging requirements
and adding in constraints from the invisible $Z$ width as measured by
LEP, $\Gamma_Z$, a general 2HDM parameter scan has been performed in
the range $1 \le m_h \le 120\;\rm GeV$, $3\;\rm GeV \le m_A \le 2\;\rm
TeV$, $0.4 \le \tan \beta \le 58.0$ and for $\alpha = \pi/2, \pi/4, 0,
-\pi/4, -\pi/2$.

\begin{figure}[ht]
\psfig{figure=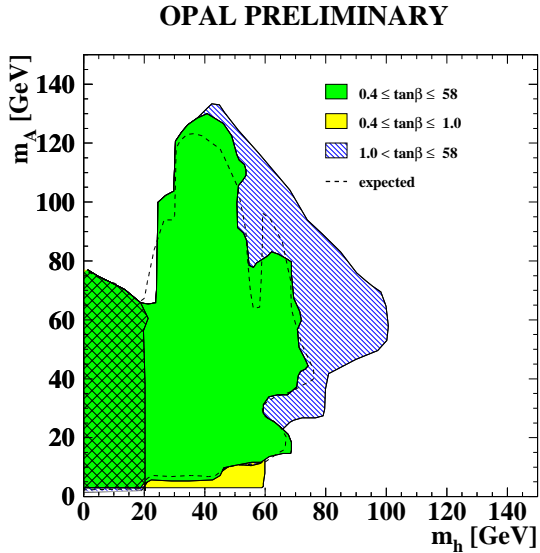,width=0.49\textwidth,clip=}
\hspace*{0.00\textwidth}
\parbox[]{0.50\textwidth}{
\vspace*{-3.50cm}
\caption[t]{
         \label{fig:opal_2hdm} Excluded $(m_A, m_h)$ region
         independent of $\alpha$, together with the expected exclusion
         limit. A particular $(m_A, m_h)$ point is excluded at 95\% CL
         if it is excluded for $0.4 \le \tan \beta \le 58.0$ (dark
         grey region), $0.4 \le \tan \beta \le 1.0$ (lighter grey
         region) and $1.0 \le \tan \beta \le 58.0$ (hatched region)
         for $-\pi/2 \le \alpha \le \pi/2$. The cross-hatched region
         is excluded using the constraints from $\Gamma_Z$
         only. Expected exclusion limits are shown as a dashed line.}}
\end{figure}

The exclusion limits are presented in several projections of the
parameters involved, where a point is excluded at 95\% confidence
level if it is excluded for all parameters scanned over. An example
for such a projection of excluded parameter regions is shown in
figure~\ref{fig:opal_2hdm}. Further details of the scan and its
results can be found in \cite{opal_2hdm_scan}. A combination of all
available LEP data in such a general 2HDM (type II) interpretation is
planned for the near future.

\subsection{MSSM Higgs Searches}
The Higgs boson sector of the MSSM corresponds to a 2HDM(type-II)
model. This implies that the decay branching ratios of the Higgs
bosons to fermions depend not only on the masses, but also on the
values of $\alpha$ and $\beta$ while the production cross-sections are
related to the SM ones as detailed in equation~\ref{eqn:2hdm_1}
and~\ref{eqn:2hdm_2}. For this analysis, the searches for the $e^+e^-
\rightarrow h^0Z^0$ processes which are used in the Standard Model
interpretations, are combined with the searches for the $e^+ e^-
\rightarrow h^0 A^0$ process. The $b\bar{b}$ and $\tau^+\tau^-$ decays
of the $h^0$ and $A^0$ are dominant for such models, and the searches
concentrate on these decays only. In addition, for models in which the
decay branching ratios of the Higgs to $b\bar{b}$ and $\tau^+\tau^-$
are suppressed, the flavour-independent results of
section~\ref{sec:flavindep} are also used. The $h^0 Z^0$ and $h^0 A^0$
searches at LEP shown here use a combination of all LEP Higgs searches
conducted at centre-of-mass energies between $\sim 88\;\rm GeV$ and
209 GeV.

The presence of an MSSM Higgs boson is tested using three benchmark
scenarios. (a) The {\it no-mixing} scenario assumes no
mixing between the left- and the right-handed stop quarks, where the
following parameters are chose: $M_{SUSY} = 1\;\rm TeV/c^2,\, M_2 =
200\;\rm GeV/c^2,\, \mu = -200\;\rm GeV/c^2,\, X_t (=A - \mu \cot \beta) =
0,\, 0.4 < \tan \beta < 50$ and $4\;\rm GeV/c^2 < m_A < 1\;\rm TeV/c^2$
with $m_{top} = 175\;\rm GeV/c^2$.  (b) The {\it $m_h$-max} scenario
is designed to yield the maximal value of $m_h$ in the model. It
therefore corresponds to the most conservative range of $\tan
\beta$-values for fixed values of the mass of the top quark and
$M_{SUSY}$. The choice of parameters is identical to the one in (a)
except that $X_t = 2\, M_{SUSY}$. (c) The {\it large-$\mu$} scenario
with parameters $M_{SUSY} = 400\;\rm GeV/c^2,
\mu = 1\;\rm TeV/c^2, M_2 = 400\;\rm GeV/c^2, m_{\tilde{g}} = 200\;\rm
GeV/c^2, 4 \le m_A \le 400\;\rm GeV/c^2, X_t = -300\;\rm GeV/c^2$, is
designed to illustrate choices of MSSM parameters for which the Higgs
boson $h$ does not decay into $\rm b\bar{b}$ due to large corrections
from SUSY loop processes. While kinematically fully accessible, this
scenario requires the flavour-independent search.

\begin{figure}[ht]
         \centerline{\psfig{figure=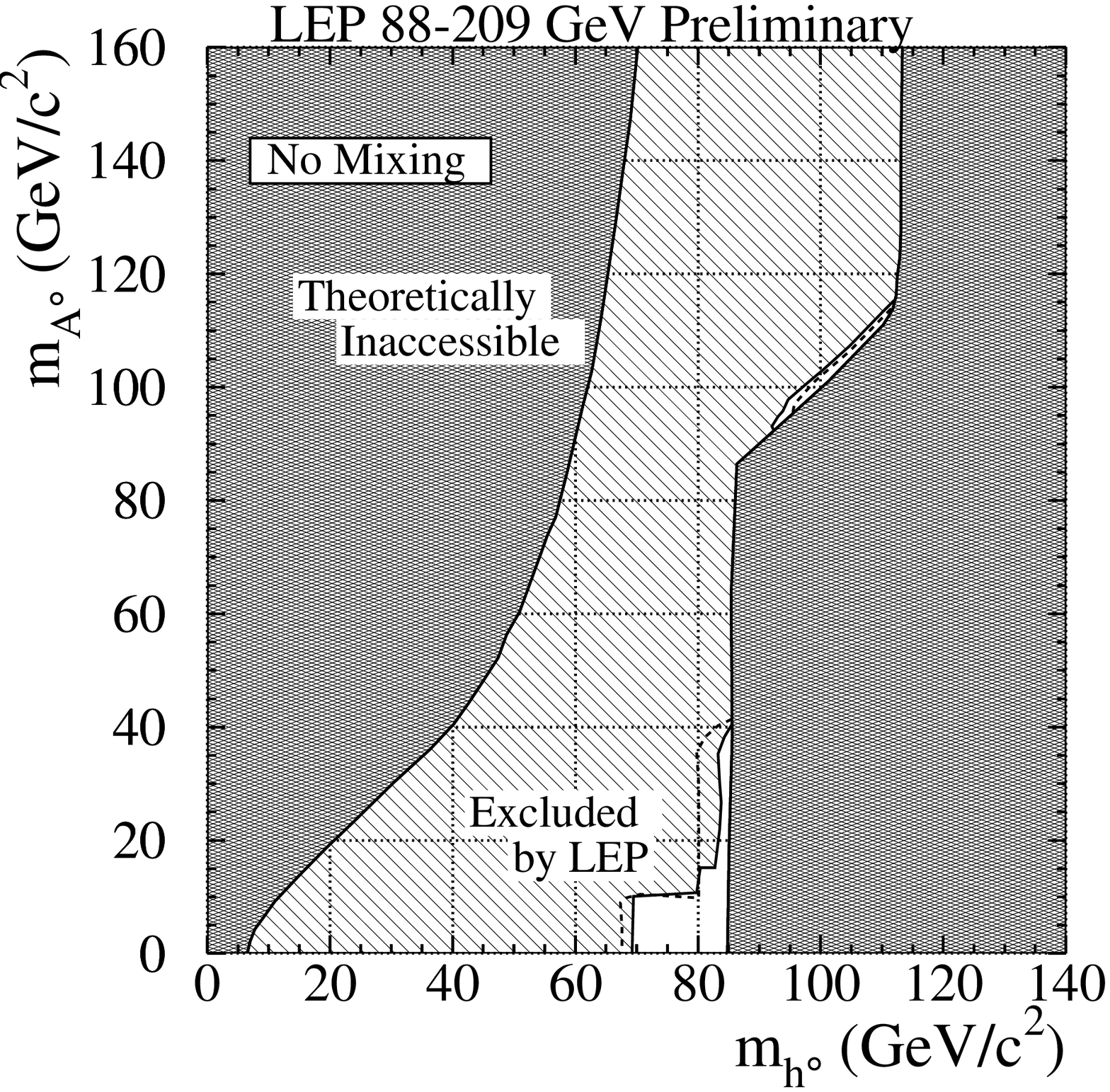,width=0.49\textwidth,clip=}
         \hfill
         \psfig{figure=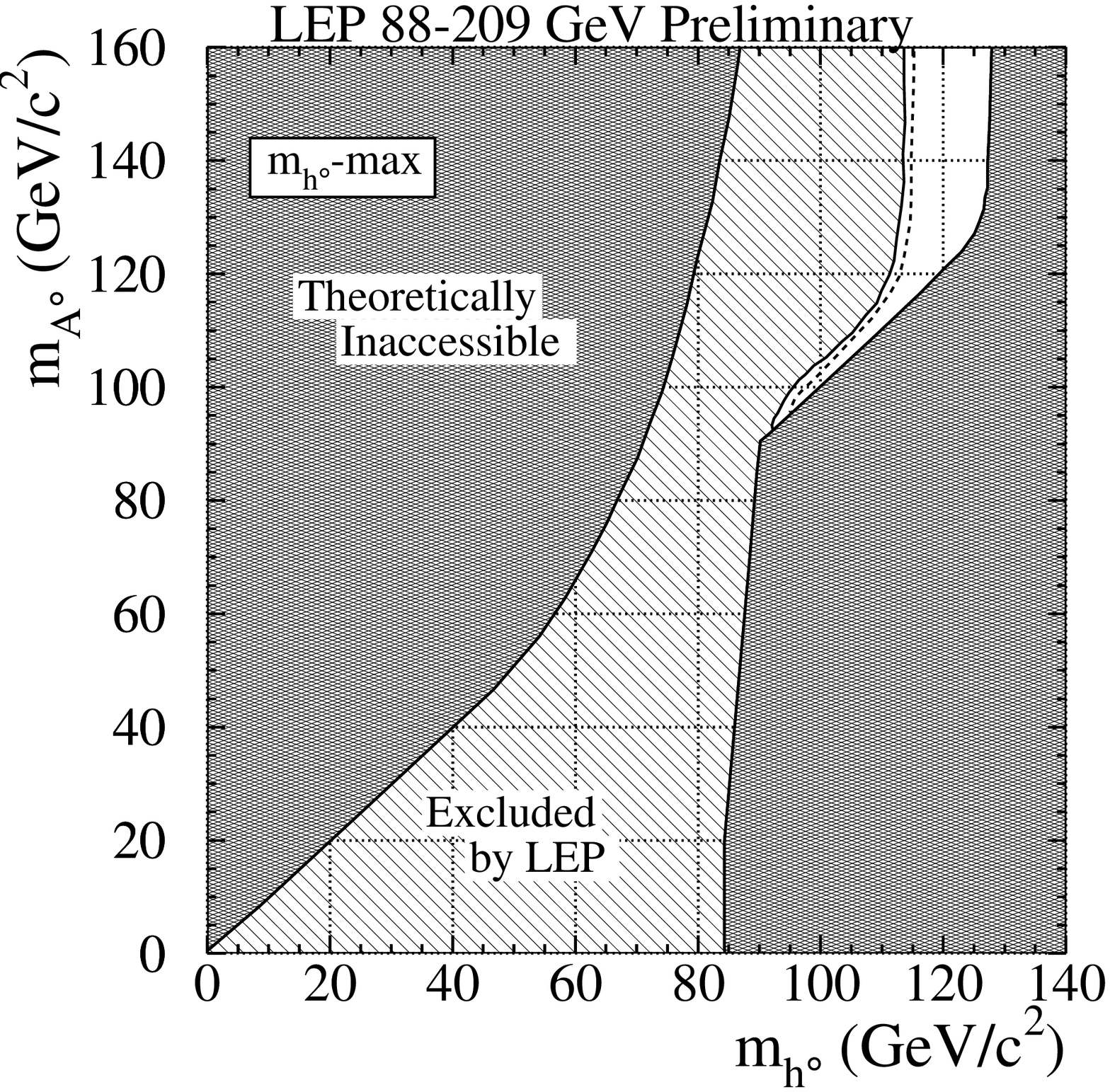,width=0.49\textwidth,clip=}}
         \caption{\label{fig:lhwg_mssm_1} The MSSM exclusion for the
         no-mixing benchmark scenario (left) and the $m_h$-max
         scenario (right). The figures show the excluded and the
         theoretically disallowed regions as a function of the Higgs
         boson masses.}
\end{figure}
Studying the {\it $m_h$-max scenario}, a few small excesses of events
are seen in single channels or single experiments at $(m_h, m_A) \sim
(83, 83) \;\rm GeV/c^2$, near $(m_h, m_a) \sim (93, 93)\;\rm GeV/c^2$,
at $m_h \approx 97\;\rm GeV/c^2$ and at about $m_h \approx 115\;\rm
GeV$. The latter is due to the events which already caused the excess
in the SM Higgs search. The significance of these excesses is always
close to 2$\sigma$ and can, in the case close to the mass diagonal, be
excluded by the combination of channels and experiments.
Figure~\ref{fig:lhwg_mssm_1} shows the excluded region for scenario
(a) and (b) in the Higgs mass plane. The {\it no-mixing} scenario is
almost completely excluded and further constraints in the low $m_A$
region can be imposed using the searches for the charged Higgs
bosons. The {\it $m_h$-max} scenario shows unexcluded region at large
$m_h$ and large $\tan \beta$.  The corresponding limits in Higgs
masses and $\tan
\beta$ are given in table~\ref{tab:lhwg_mssm_1}.
\begin{table}[ht]
\begin{tabular}{|c|c|c|c|} \hline
Scenario & $m_h$ limit $(\rm GeV/c^2)$ & $m_A$ limit $(\rm GeV/c^2)$ &
Excluded $\tan \beta$ \\
         & obs. (exp.) & obs. (exp.) & observed limit (expected limit) \\ \hline
$m_h$-max & 91.0 (94.6) & 91.9 (95.0) & $\rm 0.5 < \tan \beta < 2.4 \, (0.5 < \tan \beta < 2.6) $ \\
no-mixing & 91.5 (95.0) & 92.2 (95.3) & $\rm 0.7 < \tan \beta < 10.5 \, (0.8 < \tan \beta < 16.0)$ \\ \hline
\end{tabular}
\caption{\label{tab:lhwg_mssm_1}
Limits on $m_h$ and $m_A$ in the $m_h$-max and no-mixing benchmark
scenarios for $m_{top} = 174.3\;\rm GeV$. The median expected limits
in an ensemble of SM background-only experiments are listed in
parentheses.}
\end{table}

Adding in the flavour-independent analyses provides enough sensitivity
to exclude the entire {\it large-$\mu$} scenario at 95\% confidence
level, also the previously unexcluded parameter points. 
Further details on the data, the analyses and the model
interpretations can be found in \cite{lhwg_mssm}.

\subsection{Charged Higgs Searches}
2HDM models also predict charged Higgs bosons. At LEP2 energies
charged Higgs bosons are expected to be produced mainly through the
process $e^+e^- \rightarrow H^+H^-$. 2HDMs do not predict the $H^\pm$
mass, and the tree-level cross-section is fully determined by the
mass. The searches are carried out under the assumption that the two
decays $H^+ \rightarrow c\bar{s}$ and $H^+ \rightarrow \tau^+
\nu$ exhaust the $H^+$ decay width, but the relative branching ratio
is free.

The four LEP experiments use for this search in total $2500\;\rm
pb^{-1}$, $510\;\rm pb^{-1}$ of which above 206 GeV centre-of-mass
energy. As described in \cite{lhwg_chh}, the search channels are:
$(c\bar{s})(\bar{c}s), (\tau^+ \nu)(\tau^- \bar{\nu})$ and the mixed
mode $(c\bar{s})(\tau^-\bar{\nu}) +
(\bar{c}s)(\tau^+\nu)$. 

Taking the lowest observed limit for any branching ratios, the charged
Higgs boson can be excluded up to $m_{H^\pm} = 78.6\;\rm GeV/c^2$
while the corresponding expected limit is $78.8\;\rm GeV/c^2$ at 95\%
CL. The observed upper limit on the charged Higgs cross-section as a
function of the charged Higgs boson mass is found to be always within
the 1 $\sigma$ band around the expected median, except a small range
around $m_{H^\pm} \approx 67\;\rm GeV/c^2$ reaching about 2 $\sigma$,
caused by an excess of events in one experiment (L3). This effect is
under investigation.

In the 2HDM (type I) the bosonic decay $H^\pm \rightarrow {W^\pm}^*
A^0$ dominates near $\tan \beta >1$, if kinematically
allowed. Recently a new analysis \cite{opal_pn472}, searching for $H^+
H^- \rightarrow W^*A\, W^*A \rightarrow q\bar{q}b\bar{b}\,
q\bar{q}b\bar{b}$, $H^+ H^-
\rightarrow W^*A\, W^*A \rightarrow l \nu b\bar{b}\, q\bar{q}b\bar{b}$,
$H^+ H^- \rightarrow \tau \nu \, W^*A \tau \nu\, q\bar{q}b\bar{b}$ has
been performed.  The exclusion limits range from $m_{H^\pm} > 60\;\rm
GeV/c^2$ to $m_{H^\pm} > 80\;\rm GeV/c^2$ and from $m_A > 10\;\rm
GeV/c^2$ to $m_A > 72\;\rm GeV/c^2$, depending on $\tan \beta$ and
$(m_{H^\pm}, m_A)$.  
As soon as more than one experiment has carried out such an analysis
the results may be combined by the LEP Higgs Working Group.

\subsection{Fermiophobic Higgs Searches}
In 2HDM (type I) and other models, the Higgs coupling to fermions can
be small and the Higgs bosons therefore decay preferentially to pairs
of bosons. These are the so-called ``fermiophobic'' Higgs
bosons. Inspired by several fermiophobic models, a benchmark
fermiophobic model is defined, with Standard Model production rates
and channels, but with the fermionic channels closed. Since the
photonic Higgs decays dominate in this scenario below $m_H \approx
90\;\rm GeV$, the LEP experiments performed searches for $hZ$
production with subsequent $h \rightarrow \gamma \gamma$ decays
\cite{lhwg_gaga}.

The combination of the LEP data results in an upper limit on the
branching ratio $B(h \rightarrow \gamma \gamma)$ as a function of the
Higgs bosons mass $m_h$. For the fermiophobic benchmark a lower Higgs
mass limit of $m_h > 108.2\;\rm GeV/c^2$ with an expected median limit
of $m_h > 109.0\;\rm GeV/c^2$ is obtained.

DELPHI and L3 have turned their photonic Higgs searches also into
limits on anomalous couplings on $H \rightarrow \gamma \gamma$, $H
\rightarrow \gamma Z$ and $H \rightarrow ZZ$
\cite{DELPHI_L3_anomalous} which might be combined in the future.

\subsection{Search for Light Yukawa Production}
Within a 2HDM a search for the Yukawa process $e^+e^- \rightarrow bb
A/h \rightarrow b\bar{b}\tau^+\tau^-$ in the mass range of 4-12 GeV
has been performed, using the data collected by the OPAL
\cite{opal_light_yukawa}. Within a CP-conserving 2HDM(type II) the
cross-section for Yukawa production depends on $\xi_d^A = |\tan
\beta|$ and $\xi_d^h = | \sin \alpha / \cos
\beta|$ for the production of the CP-odd $A$ and the CP-even $h$,
respectively. From the data 95\% CL upper limits are derived for
$\xi_d^A$ within the range of 8.5 to 13.6 and for $\xi_d^h$ between
8.2 to 13.7, depending on the mass of the Higgs boson, assuming a
branching ratio fraction into $\tau^+ \tau^-$ of 100\%. An
interpretation of the limits within a 2HDM (type II) with Standard
Model particle content is also given.


\section{Search for Doubly Charged Higgs Bosons}
Doubly charged Higgs bosons $(H^{\pm\pm})$ appear in theories beyond
the Standard Model, for example in left-right symmetric models. Such
models in which $SU(2)_R$ gauge symmetry is broken by triplet Higgs
fields do not conserve baryon and lepton numbers. Searches for
pair-produced doubly charged Higgs bosons have recently been performed
by DELPHI and OPAL \cite{delphi_opal_doublych} using the data
collected at centre-of-mass energies between 189 and 209 GeV. At the
95\% CL, lower limits for the mass of doubly charged Higgs bosons in
left-right symmetric models have been set at $99.1\;\rm GeV/c^2$ and
$98.5\;\rm GeV/c^2$ by DELPHI and OPAL, respectively. These results
might be combined by the LEP Higgs working group in the future.

\section{Conclusion}
The four LEP experiments have searched for Higgs bosons in several
models with varying model dependence. Most of the results have been
combined by the LEP Higgs Working Group resulting in various exclusion
limits on Higgs boson masses, production cross-sections or decay
branching ratios. In the search for the Standard Model Higgs, a
$\approx 2\, \sigma$ excess has been observed at a mass of $m_h =
115.6\;\rm GeV/c^2$. First final combinations by the LEP Higgs Working
Group are planned for the summer 2002.

\section*{Acknowledgments}
I would like to thank the organisers, in particular Jean Tran Thanh
Van, for this and all the previous `Rencontres de Moriond' conferences
and Peter Igo-Kemenes for careful reading of this manuscript.

\end{document}